\documentclass[iop,apjl,twoside]{emulateapj}  

\newcommand{\Note}[1]{\hfill\parbox{4in}{\footnotesize\bfseries{\color{red}\texttt{[#1]}}}\\}
\renewcommand{\Note}[1]{\relax}  

\newcommand{\Change}[1]{{\bf#1}}
\renewcommand{\Change}[1]{#1}  

\usepackage{xspace}
\usepackage{amssymb}
\usepackage{times}
\usepackage{verbatim}
\usepackage{color}

\newcommand{\cmmthree}{\,\mathrm{cm\mthree}}

\newcommand{\kev}   {\,\mathrm{keV}}
\newcommand{\kms}   {\,\mathrm{km\,s\mone}}
\newcommand{\ks}    {\,\mathrm{ks}}

\newcommand{\mang}  {\,\mathrm{{\AA}}\xspace}
\newcommand{\mk}    {\,\mathrm{MK}}

\newcommand{\mone}  {^{-1}}

\newcommand{\mthree}{^{-3}}

\newcommand{\acis}  {{ACIS}\xspace}

\newcommand{\chan}  {{\it Chandra}\xspace}
\newcommand{\suz}  {{\it Suzaku}\xspace}
\newcommand{\ciao}  {{CIAO}\xspace}

\newcommand{\hetg}  {{HETG}\xspace}

\newcommand{\letg}  {{LETG}\xspace}

\newcommand{\mucol} {{$\mu\,$Col}\xspace}
\newcommand{\sigori} {{$\sigma\,$Ori}\xspace}

\newcounter{ion} \newcommand{\eli}[2]{\setcounter{ion}{#2}#1{~\sc\roman{ion}}}

\shorttitle{The Massive Hot Wind of \mucol}
\shortauthors{Huenemoerder et al.}

\begin{document}

\title{On the Weak-Wind Problem in Massive Stars: X-ray Spectra Reveal
  a Massive Hot Wind in $\mu\,$Columbae}

\author{David P.\ Huenemoerder\altaffilmark{a},
Lidia M. Oskinova\altaffilmark{b},
Richard Ignace\altaffilmark{c},
Wayne L. Waldron\altaffilmark{d},
Helge Todt\altaffilmark{b},
Kenji Hamaguchi\altaffilmark{e,f},
Shunji Kitamoto\altaffilmark{g}
\ \\
}

\altaffiltext{a} {Massachusetts Institute of Technology, Kavli
  Institute for Astrophysics and Space Research, 70 Vassar St.,
  Cambridge, MA, 02139, USA}

\altaffiltext{b} {Institute for Physics and Astronomy, University of
  Potsdam, 14476 Potsdam, Germany}

\altaffiltext{c} {Department of Physics and Astronomy, East Tennessee
  State University, Johnson City, TN 37614, USA}

\altaffiltext{d}{Eureka Scientific Inc., 2452 Dellmer St., Suite 100, Oakland, CA 94602, USA}   

\altaffiltext{e}{CRESST and X-ray Astrophysics Laboratory NASA/GSFC,
  Greenbelt, MD 20771, USA}

\altaffiltext{f}{Department of Physics, University of Maryland,
  Baltimore County, 1000 Hilltop Circle, Baltimore, MD 21250, USA} 


\altaffiltext{g}{Department of Physics, Rikkyo University, Tokyo
  171-8501, Japan}   



\begin{abstract}

  $\mu\,$Columbae is a prototypical weak-wind O-star for which we have
  obtained a high-resolution X-ray spectrum with the \chan\
  \letg/\acis instrument and a low resolution spectrum with {\em
    \suz}.  This allows us, for the first time, to investigate the
  role of X-rays on the wind structure in a {\it bona fide} weak-wind
  system and to determine whether there actually is a massive, hot
  wind.  The X-ray emission measure indicates that the outflow is an
  order of magnitude greater than that derived from UV lines and is
  commensurate with the nominal wind-luminosity relationship for
  O-stars.  Therefore, the ``weak-wind problem''---identified from
  cool wind UV/optical spectra---is largely resolved by accounting for
  the hot wind seen in X-rays.  From X-ray line profiles, Doppler
  shifts, and relative strengths, we find that this weak-wind star is
  typical of other late O dwarfs.  The X-ray spectra do not suggest a
  magnetically confined plasma---the spectrum is soft and lines are
  broadened; \suz spectra confirm the lack of emission above 2 keV.
  Nor do the relative line shifts and widths suggest any wind
  decoupling by ions.  The He-like triplets indicate that the bulk of
  the X-ray emission is formed rather close to the star, within 5
  stellar radii. Our results challenge the idea that some OB stars are
  ``weak-wind'' stars that deviate from the standard wind-luminosity
  relationship. The wind is not weak, but it is hot and its bulk is
  only detectable in X-rays.

\end{abstract}




\section{Introduction}

The outflow of stellar winds from massive OB-type stars is an
important process which affects both the chemical enrichment and
kinetics of the interstellar medium
\citep[e.g.][]{Leitherer:Robert:Drissen:1992}.  The mass-loss itself
is enough to change the evolution of the star, which ends its life
in a supernova explosion, also profoundly changing its environment.
Hence, quantitative understanding massive star winds is important not
only as a basic component of stellar astrophysics, but also for
understanding cosmic feedback on galactic scales throughout cosmic
history.

While some basic physics of stellar winds in massive stars is well
established---that winds are accelerated by photoelectric absorption
of the intense ultraviolet radiation field by a multitude of metal
lines, that an instability can lead to wind-shocks which generate
X-rays---there are still puzzles to be solved.  One of these is the
``weak-wind'' problem, in which UV line diagnostics clearly show a
wind signature in classical P~Cygni line profiles, but modeled mass
loss rates can be discrepant by more than an order of magnitude from
values expected based on O-star statistical trends and theoretical
foundations, specifically the wind momentum-luminosity relation
\citep[e.g.][]{Puls:Kudritzki:Herrero:al:1996}.  Factors of a few in
mass loss are enough to be significant for stellar evolution and
cosmic feedback \citep[e.g.][]{Puls:Vink:Najarro:2008}.

It has long been known that photoionization by X-rays can alter the
ionization balance in the wind regions where the UV lines are formed
\citep{Waldron:1984,Macfarlane:Waldron:al:1993}.  There is a
theoretical degeneracy in that different values of mass loss rate
($\dot M$) and X-ray luminosity ($L_x$) can produce very similar UV
line profiles \citep{Puls:Vink:Najarro:2008,
  Marcolino:Bouret:al:2009}; direct knowledge of the X-ray spectrum is
thus important for reliable determination of wind parameters.  Another
possibility is that cool and hot plasma emission originate from
different volumes or densities; that is, clumping can affect the
interpretation \citep{Hamann:Feldmeier:Oskinova:2008}.  \mucol belongs
to the weak-wind domain defined by \citet{Lucy:2010} in which a star's
rate of mechanical energy loss in a radiatively driven wind is less
than the radiative output from nuclear burning; Lucy showed
that there is a huge disparity between the theoretically expected
$\dot M$ and values derived from UV and optical spectra.
\citet{Lucy:2012} developed a phenomenological model and suggested
that in low-luminosity O-type stars, the volumetric roles of hot and
cool gas are possibly reversed compared to O-type stars of high
luminosity; thus in the weak-wind stars, a larger volume is occupied
by the hot gas than by the cool gas.

\mucol (HD~38666) is an O9.5~V runaway (and single) star and is one of
the weakest wind Galactic O-type stars \citep[e.g., see Figs.\ 39 and
41 in][]{Martins:Schaerer:al:2005}.  These factors are what motivated
our spectroscopic study of the prototypical weak-wind system, \mucol,
at high-resolution with \chan and at low resolution but greater
sensitivity at higher energies with \suz.  In this letter, we
concentrate on the primary empirical results from the X-ray spectral
analysis of \mucol.  A subsequent paper will investigate the influence
of the X-rays on the cool wind component (Todt et. al, in
preparation).

%
\begin{figure*}[!htb]
  \centering\leavevmode
  \includegraphics[width=0.98\textwidth,viewport=1 1 530 140]{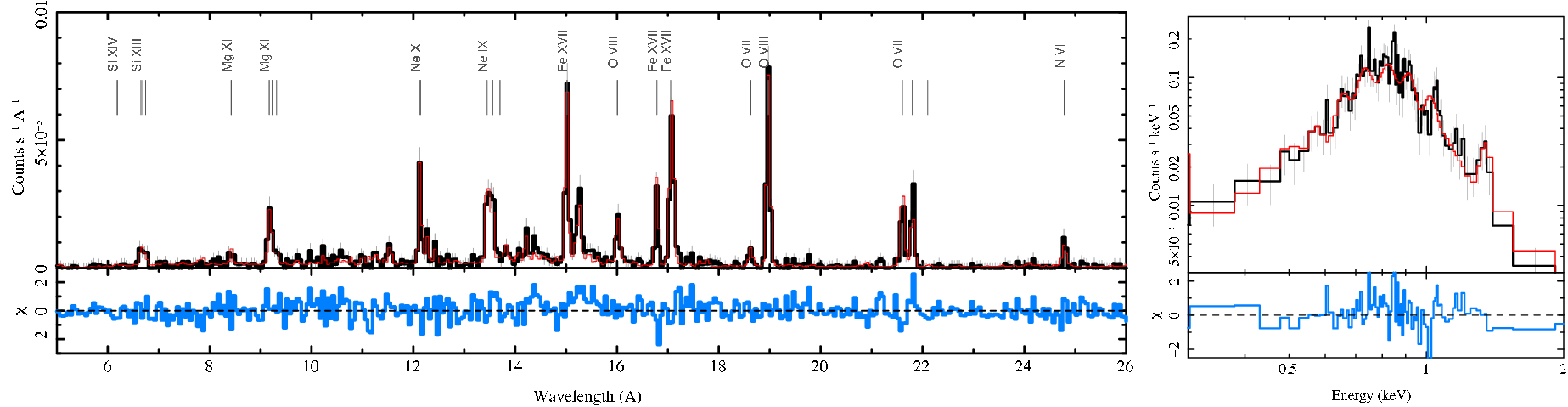}
  \caption{\mucol\ \chan/\letg/\acis spectrum (left); black: observed
    count-rate; thin-gray line (red in the online version): model; the
    lower panel shows the $\chi^2$ residuals against a broken powerlaw
    APEC model modified for photoexcitation of triplets.  Lines are
    broader than the instrumental width. The \suz spectrum is shown on
    the right (black).  For conciseness, we have summed the counts
    from the three detectors; such is not recommended when fitting,
    but it provides a good summary visualization of the data.  The
    folded \chan-derived model is in gray (red in the online version),
    and residuals below.  We emphasize that the \chan-derived model
    was not fit to the \suz spectrum, only folded through the response
    to provide model counts and residuals; the model agrees very well
    without any adjustments.  It is significant that there is little
    or no flux detected above $2\kev$ where \suz has substantial
    sensitivity.}
  \label{fig:mucolspec}
\end{figure*}

\begin{deluxetable}{ll|ll}
\tablecolumns{4}
\tablewidth{0.95\columnwidth}
\tablecaption{\mucol Properties}
\tablehead{
  \colhead{Property}&  \colhead{Value} &\colhead{Property}&  \colhead{Value}
}
\startdata
Spectral Type&
O9.5 V&
$d\,$[pc]\tablenotemark{a}&
408\\

$T_\mathrm{eff}\,$[K]\tablenotemark{b}&
$33000$&
$R/R_\odot$\tablenotemark{b,e}&
$4.6$\\

$\log(L_\mathrm{bol}/L_\odot)$\tablenotemark{b,e}&
4.4&
$v_\infty[\kms]$\tablenotemark{b}&
$1200$\\

$\dot M_{uv}\,\mathrm{[M_\odot/yr]}$\tablenotemark{b}&
$10^{-9.5}$&
$N_\mathrm{H}\,[\mathrm{cm^{-2}}]$\tablenotemark{c}&
$5\times10^{19}$\\

$f_x(1-40\mang)[\mathrm{cgs}]$\tablenotemark{d,f}&
$6\times10^{-13}$&
$\log(L_\mathrm{x}/L_\mathrm{bol})$\tablenotemark{d}&
$-6.9$\\
$EM_x [\cmmthree]$\tablenotemark{d}&
$10^{54.1}$&
$T_\mathrm{max}[\mk]$\tablenotemark{d}&
$4.4$\\
$v_{\infty,x}[\kms]$\tablenotemark{d}& $1600\,(\pm275)$
&&\\ 
$f/i(${\eli{O}{7}}$)$\tablenotemark{g}&
$<0.01$ &
$r($\eli{O}{7}$)$\tablenotemark{g}&
$<3.3$\\
$f/i(${\eli{Ne}{9}}$)$\tablenotemark{g}&
$0.04$--$0.14$ &
$r($\eli{Ne}{9}$)$\tablenotemark{g}&
$2.3$--$4.4$\\
$f/i(${\eli{Mg}{11}}$)$\tablenotemark{g}&
$>0.2$ &
$r($\eli{Mg}{11}$)$\tablenotemark{g}&
$>2.1$
\enddata
\tablenotetext{a}{From the Hipparcos parallax, as re-evaluated by \citet{vanLeeuwen:2007}}
\tablenotetext{b}{\citet{Martins:Schaerer:al:2005}}
\tablenotetext{c}{\citet{Cassinelli:al:1994,Howk:Savage:Fabian:1999}}
\tablenotetext{d}{This work}
\tablenotetext{e}{Adjusted for the adopted distance.}
\tablenotetext{f}{The flux is as observed at Earth, with foreground absorption.}
\tablenotetext{g}{$f/i$ gives the ratio of the forbidden to
  intercombination line fluxes, and $r$ is the radius of formation in
  units of the stellar radius, as derived from PoWR models.}
\label{tbl:props}
\end{deluxetable}
\section{Analysis}

We have obtained a $232\ks$ exposure of \mucol with the \chan\
\letg/\acis instrument (observation IDs
\dataset[ADS/Sa.CXO#obs/12349]{12349},
\dataset[ADS/Sa.CXO#obs/12349]{12350},
\dataset[ADS/Sa.CXO#obs/12349]{13422}, PI L.\
Oskinova). Figure~\ref{fig:mucolspec} shows the count-rate spectrum.
Since \mucol is a single star, there are no ambiguities present as
when interpreting observations of binary systems with composite
spectra or colliding wind emission.  Relevant stellar properties are
given in Table~\ref{tbl:props}.

There are several key X-ray spectral stellar wind diagnostics.  The
line profile is sensitive to the wind opacity and velocity field
\citep{Macfarlane:al:1991,Owocki:Cohen:2001}; the line centroid and
width are useful proxies, being sensitive to wind parameters governing
the detailed line shape.  Emission line strengths are indicative of
plasma temperatures and elemental abundances.  The continuum at the
shortest wavelengths available ($2$--$10\mang$, $1$--$6\kev$) is also
very sensitive to the highest temperatures present.  The He-like
triplets are sensitive to electron density and the UV radiation field
through collisional and photoelectric excitation which can depopulate
the forbidden-line level, weakening it while strengthening the
intercombination lines \citep{Gabriel:69,Blumenthal:Drake:al:1972}; in
O-stars, the UV field typically dominates the depopulation and hence
He-like line ratios are diagnostic of radius of formation
\citep{Waldron:Cassinelli:2001}.  Very close to the photosphere,
density effects could also become significant.

\begin{figure}[!htb]
  \centering\leavevmode
  \includegraphics[width=0.85\columnwidth]{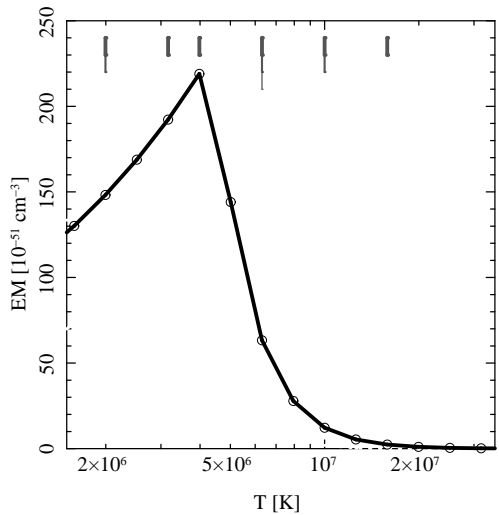}
  \caption{
    The best fit emission measure distribution is given by the
    thick solid line.  Circles mark the 1 dex temperature intervals,
    the resolution of the emmissivity database, and the value plotted
    is the emission measure integrated over 0.1 dex.  Vertical bars
    near the top mark the temperatures of maximum emissivity for
    detected emission lines, and the abscissa's range spans the
    temperatures where these lines have greater then 50\% of their
    maximum emissivity.  The instruments have significant sensitivity
    from 3--6 \AA\ such that EM above $10^7\,$K would be apparent in
    lines and continuum. 
  }
  \label{fig:emd}
\end{figure}

We have fit a global model to the X-ray spectrum, using standard
products produced by \ciao \citep[version 4.3, and associated
calibration database]{CIAO:2006}, using ISIS \citep{Houck:00} and the
collisional ionization equilibrium emissivities in AtomDB
\citep[version 2.0,][]{Smith:Brickhouse:2008}.  For the plasma model,
we used a broken powerlaw emission measure distribution \Change{(EMD;
  which in differential form is defined as $n_e n_h dV/dT$)}, with
variable abundances for significant ions, a Doppler shift, and line
profiles defined by a global Gaussian Doppler broadening term.  Our
models show that the absorption of X-rays in the cool stellar wind is
negligible, and spectra can be well fitted neglecting wind absorption.
Abundances, velocity, and broadening were common over all temperature
components.  The resulting fit is shown as the red curve in
Figure~\ref{fig:mucolspec}.
  The emission measure distribution is shown in Figure~\ref{fig:emd}.

Abundances (referenced to solar photospheric values of
  \citet{Asplund:Grevesse:al:2009}) were about half of solar (O, Si,
  and Fe) or near solar (N, Ne)---we could not obtain a good fit using
  abundances all set to solar.  This may be due to the adopted smooth
  functional form of the emission measure distribution, since the EMD
  and abundances are somewhat degenerate.  Trial fits with discrete
  temperature components were also poorer when using solar abundances.
  To explore this somewhat further, since the integrated emission
  measure is fundamental to our main result, we evaluated the 90\%
  confidence limits of the model normalization, froze this at each of
  the high and low limits, and re-fit the spectrum to obtain new EMD
  and abundances.  We could have half the integrated EM with relative
  abundances increased to 0.7--1.5, or we could have about triple the
  best-fit EM, also with slightly modified abundances.  (We note that
from UV spectra, \citet{Fitzpatrick:Massa:1999} also found low
abundances for \mucol, but did not believe their results plausible and
attributed them to model deficiencies.)

The broken powerlaw model is empirically justified in that it
  provides a necessary multi-thermal model and does fit well with
  relatively few parameters.  Such an EMD can be physically justified
  by hydrodynamic shock models which predict a wide range of
  temperatures over a large range of radii
  \citep{Feldmeier:Puls:Pauldrach:1997}.  Other OB-stars have also
  shown similar, empirically determined EMD
  \citep{Wojdowski:Schulz:2005}.  The important point here is to
  obtain an order-of-magnitude estimate of the EM, and that can be
  done with a variety of plausible models.  Also given that the X-ray
  emitting plasma is optically thin, it does not matter (for the EMD)
  where the emission originates, under the assumption that abundances
  are independent of temperature and density.

Model properties are given in Table~\ref{tbl:props}; the volume
emission measure, $EM_x$, is integrated over the temperature
range from 1--100 MK.

We have observed \mucol using the \suz satellite for $26\ks$
(\dataset[ADS/Sa.Suzaku#obs/405059010]{observation ID 405059010, PI
  L. Oskinova}).  We used the HEASoft 6.9 and the CALDB xis20100123
for the XIS analysis; the HXD did not detect any signal.  The \suz
spectrum is consistent with the above model, with no flux above
$2\kev$.  Figure ~\ref{fig:mucolspec} shows the \suz spectrum, the
folded \chan-derived model, and residuals; there is good agreement.

\begin{figure}[!htb]
  \centering\leavevmode
  \includegraphics[width=0.85\columnwidth,viewport=25 30 500 470]{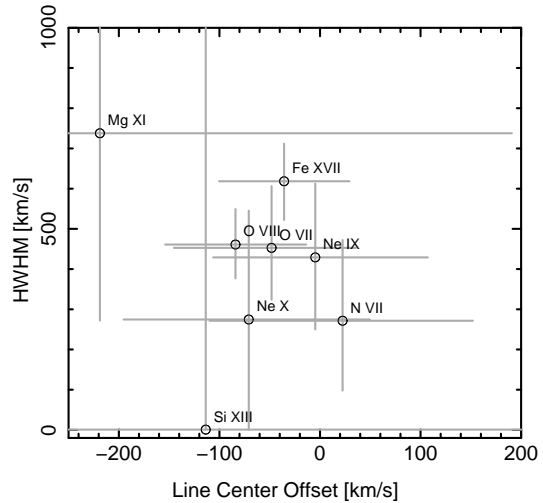}
  \caption{\mucol\ emission line Doppler offset and width; errorbars
    give 90\% confidence limits.  Due to lower resolution at
    shorter wavelengths, the \eli{Mg}{11} and \eli{Si}{13} lines are
    poorly constrained.  The systemic line-of-sight velocity of
    $109.2\kms$ \citep{Evans:1967} has been removed.}
  \label{fig:mucollines}
\end{figure}

Global fits cannot necessarily provide detailed information present in
individual lines, particularly if lines from different ions have
different characteristics from local conditions (e.g., velocity and
temperature gradients).  Hence, we also fit the stronger lines in the
\letg/\acis spectrum with Gaussians (folded through the instrument
response), with a continuum derived from the global plasma model.  The
parameters relevant here are the centroids and widths
(Figure~\ref{fig:mucollines}).  A non-zero centroid indicates a
wind, being skewed to the blue by disk occultation of the receding
wind and by absorption in the wind.

\begin{figure}[!htb]
  \centering\leavevmode
  \includegraphics[width=0.85\columnwidth]{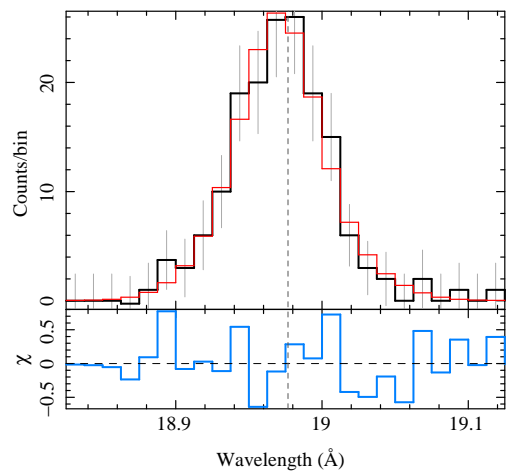}
  \caption{\mucol\ \eli{O}{8} profile (black), wind-profile
    fit (gray; red in the online version) and residuals (lower
    panel). The model has a line center offset of $24\, (\pm 54)
    \kms$, and $v_{\infty} = 1600\, (\pm 275)\kms$. The model profile
    was for $\beta=0.7$, $R_0/R_\star=1.1$, with a smooth, un-clumped wind.}
  \label{fig:oviiiwindprof}
\end{figure}

We also computed a model line shape for an expanding wind
\citep{Oskinova:al:2006} and fit this to the strongest, isolated
feature in the spectrum, \eli{O}{8} ($18.967,18.973\mang$) by
adjusting the line position, flux, and scaling the width
(Figure~\ref{fig:oviiiwindprof}).
We used an un-clumped model profile with $\beta=0.7$ and
  $R_0/R_\star=1.1$.
We find the wind is thin enough that a small asymmetry and shift
is due to the stellar disk occultation of the receding hot wind.
The blueshift in the Gaussian fit to \eli{O}{8}
of $-84\kms$ (Figure~\ref{fig:mucollines}) is consistent with
the wind profile model, whose best fit is at the expected line
position (i.e., no offset, within one standard deviation
accuracy of $50\kms$).  We did not achieve a good fit for the
UV-derived $v_{\infty}=1200\kms$, but had to increase the hot wind
velocity to $1600\, (\pm275)\kms$.  UV spectra do
not allow precise determination of $v_\infty$; published values range
from $1000$--$2000\kms$ \citep{Martins:Schaerer:al:2005}. Our PoWR
models showed that the \eli{C}{4}, \eli{Si}{4}, and \eli{N}{5} lines
can be equally well fit with $v_\infty$ of $1200\kms$ or $1600\kms$.
We can also achieve an equally good fit to \eli{O}{8} for
  $\beta=1$, but with $v_\infty = 2800\kms$.  More
  detailed analysis with more lines and a variety of model assumptions
  are required.  For a conservative approach, we prefer the lower
  $v_\infty$.

The final line measurements of use are the He-like triplet ratios.  We
have measured these for \eli{O}{7}, \eli{Ne}{9}, and \eli{Mg}{11}
(\eli{Si}{13} is too weakly exposed and significantly blended to be
useful).  The forbidden lines are much weakened:
The \eli{O}{7} forbidden line was not detected, and the
  \eli{Ne}{9} forbidden line was very weak.  In both cases, the
  intercombination lines are as strong as the resonance lines.
The forbidden-to-intercombination ratios ($f/i$) provide limits on the
radii of formation.  We applied a simple model assuming triplet
formation occurs at a single radius in the wind, using only excitation
by photospheric UV fluxes from model atmospheres such as TLUSTY or
ATLAS-9 \citep{Lanz:Hubeny:2007,Lanz:Hubeny:2003,Kurucz:1979}.  We
also used more sophisticated treatment using the PoWR code
\citep{Hamann:Grafener:2004} in which the wind model calculates
ionization structure, solves the energy equation, includes X-rays,
includes photospheric emission corrected for absorption lines, line
blanketing, non-LTE populations, stellar disk limb darkening, and the
diffuse UV radiation field.  Each of the methods predicts the
$f/i$-ratio vs radius $r$ of formation.  Values (which were nearly
identical for all methods) are given in Table~\ref{tbl:props} with
$r$, in units of the stellar radius.

There is some temperature dependence in the triplet ratios,
  but it is below 10\%, less than our measurement uncertainty.  The
  point-formation model also ignores the effect of any distributed
  emission.  If there is contribution from further out in the wind,
  our $f/i$-derived radii will be overestimates (since $f$ increases
  and $i$ decreases at larger radii---the distributed emission ratio
  would be larger than at any point at smaller radius).

We have also looked for variability.  The count rate in dispersed
photons appears constant; the rate cannnot change by more than 5--10\%
in time intervals of 1 ks or more.

\section{``Weak-Wind'' Stars Are Not Low Mass-Loss Winds}

The He-like triplets place the X-ray emission within about 2--5
stellar radii of the photosphere.  Line widths are consistent with
X-rays being formed at small radii.  We assume that the hot
plasma obeys the continuity equation, that the standard velocity law
holds for the hot plasma, namely $v_x = v_{\infty,x} (1 - b R_*/
r)^\beta$, and that the hot plasma exists only above some inner radius
$R_0 \ge R_*$.  For $v_\infty = 1200\kms$, $b = 0.97$
(a value based on the ratio of the sound speed at $R_\star$ to
  $v_\infty$, which provides a non-zero wind velocity, but which is
  somewhat uncertain.)  and $\beta = 1$, we expect wind velocities in
  this region to be about 800--1300$\kms$.
  Figure~\ref{fig:mucollines} shows the widths to be somewhat smaller
  at 300--700$\kms$.  This suggests that much of the X-ray emission
  originates very close to the star within the radiation-driven wind
  acceleration zone, or that the hot plasma does not follow the
  typical velocity law.  The hot plasma, however, does expand with
  relatively high velocity since the X-ray emission line profiles are
  resolved.  Detailed line-profile fitting using wind-models will be
  required to determine the structure in more detail.

With the assumptions above and the definition of the X-ray emission
measure (and $\beta=1$ to allow an analytic integration), we can
derive a simple expression for the X-ray inferred mass loss rate in
solar masses per year, also assuming that the hot wind is unclumped,
and that the cool wind is optically thin to X-rays (as justified by
the line profiles):
\begin{equation}
  \label{eq:mdot}
  \dot M_x = 3\times10^{-9}\,
  {v_{\infty,x}}
  \left[{\frac{R_\star}{R_\odot}\,
    {EM_x}\,
    \left(\frac{R_0}{R_\star}-b \right) }
  \right]^{1/2}
\end{equation}
in which $v_{\infty,x}$ is the hot wind's terminal velocity as
determined from X-ray emission line profiles (in units of $1000\kms$),
$R_\star$ is the stellar radius, $EM_x$ is the emission measure of the
hot plasma (in units of $10^{54}\cmmthree$), and $b$ is the unitless
parameter from the wind velocity law.  Using $R_0=R_\star$, $b=0.97$ and
values from Table~\ref{tbl:props}, we infer that $\dot M \approx
2\times10^{-9}~\mathrm{M_\odot\,yr\mone}$.  This is 6 times the value
derived from the UV by \citet{Martins:Schaerer:al:2005} (or more,
accounting for subsequent revisions in distance and stellar radius).
If the hot wind begins somewhere above the photosphere, then the
inferred mass loss rate will be even larger (e.g., about 20 times for
$R_0/R_\star=1.5$).

The above $\dot M_x$ uses our $EM_x \sim 10^{54}$. If were to assume
the much lower $\dot M_{uv}$, we would infer a much smaller emission
measure for a spherically symmetric wind.  This means that most of the
wind is hot (or denser) than the UV-emitting plasma, a situation also
noticed in the study of main sequence B stars
\citep[e.g.][]{Cassinelli:al:1994}.

We conclude that the hot wind of $\mu$~Col must have a larger volume or
greater density than the cool wind.  The wind is not weak, but it is
hot and its bulk is only detectable in X-rays.

Our observations exclude other proposed weak-wind explanations, such
as magnetically channeled wind shocks (MCWS) or frictionally decoupled
winds.  The X-ray spectrum of \mucol is not characteristic of MCWS in
which hot plasma is held in stationary structures close to the stellar
photosphere and which reach high temperatures from collision of
funneled high velocity winds
\citep{Babel:Montmerle:1997,Townsend:Owocki:uddoula:2007}. The plasma
of \mucol is not extremely hot ($T\gtrsim10\mk$) nor are lines
unshifted and unresolved.  Decoupling of ions from neutrals could
occur at low densities or very low metallicities, creating a two-fluid
system \citep[c.f.,] []{Springmann:Pauldrach:1992, Krticka:2001,
  Martins:Schaerer:al:2004}.  Frictional heating by decoupled ions
seems unlikely: the plasma is not extremely cool ($\lesssim1\mk$) with
different widths and velocities for different ions (within our
sensitivity; Fig.~\ref{fig:mucollines}).  Given the X-ray
$EM_x$ and likely radii of formation, the density is not as low as
once presumed.  The relative Fe abundance from our fits is about 0.5
solar, not low enough to cause decoupling.

\citet{Lucy:2010} reduced the weak-wind discrepancy through
theoretical arguments. The X-ray spectra of \mucol support this with
empirical evidence for a dominant hot wind, even given some
uncertainty from still poorly determined factors of clumping, wind
velocity law, and location of the wind base.  \citet{Drew:al:1994}
suggested that X-rays may increase the wind ionization at the critical
point, lowering the effective radiative acceleration and mass-loss
rate.  Our \chan observations resolve the broad X-ray emission lines,
indicating that the hot plasma expands at high velocities, perhaps
even exceeding the cool wind velocity determined from UV spectra.

Our empirical findings, reached in ignorance of work by
\citet{Lucy:2012}, are in line with his claim that in late-type
O-dwarfs the ambient wind is heated to temperatures of few MK at radii
$>1.4\,{R_*}$, with cool radiatively-driven gas being confined to
dense clumps and small volume filling factor.  Further out in the wind
in his model, cool clumps are destroyed by heat conduction from the
hot plasma, and the outflow is dominated by a hot thermal wind which
reaches a supersonic terminal velocity of $\sim 1000\kms$.

In future work, we will refine our order-of-magnitude estimates using
detailed wind models.  For a relatively thin wind, we do
not expect significant qualitative changes.  Some
observational uncertainties need to be refined:  we only have
both upper and lower limits for the formation radii of \eli{Ne}{9};
whether other species form at $5\,R_*$ or below $2\,R_*$ is important
for wind shock models.  Higher spectral resolution would be of use,
especially for the hottest He-like triplet in the spectrum,
\eli{Si}{13}, to probe the deepest layers.  Higher resolution
would better determine line shifts and widths, and
better constrain wind structure.

\citet{Najarro:Hanson:Puls:2011} recently classified \sigori AB
(HD~37468, O9.5~V~+~B0.5~V) as a weak-wind system, based on infrared
spectra,  revising its mass-loss rate downward by more than
two orders of magnitude from that in \citet{Howarth:Prinja:1989}.
This star has also been observed at high resolution in X-rays
\citep{Waldron:Cassinelli:2007, Zhekov:Palla:2007,
  Skinner:Sokal:al:2008}.  Based on these prior works and our own
analysis of the \chan/\hetg spectrum, we find that \sigori has X-ray
spectral characteristics very similar to those of \mucol.  This
further corroborates our conclusions that the weak-wind phenomenon is
due to a property of the cool plasma being a minor constituent of the
wind.

\section{Conclusions}

Our results challenge the idea that some OB stars are ``weak-wind''
stars that deviate from the standard wind-luminosity relationship.
From high-resolution X-ray spectrum of \mucol, specifically
He-like lines and the total emission measure, this star does not
appear unusual relative to other O-stars except for its weak-wind
status.  Its X-ray emission measure, line widths, and centroids are in
good agreement with the OB main sequence star results of
\citet{Waldron:Cassinelli:2007}.  The volume emission
measure of the X-ray emitting plasma must be very much larger than the
cool, UV-emitting plasma, and we believe that the weak-wind problem is
reduced or eliminated when the hot and dominant component of the wind
is taken into account.  The wind is not weak, but it is hot and its
bulk is only detectable in X-rays.

\acknowledgements Acknowledgements: Support for this work was provided
by the National Aeronautics and Space Administration through \chan
Award Numbers GO1-12017A (WLW), GO1-12017B (DPH), and GO1-12017C (RI)
issued by the \chan X-ray Observatory Center, which is operated by the
Smithsonian Astrophysical Observatory for and on behalf of the
National Aeronautics Space Administration under contract NAS8-03060.
LMO was funded by DLR grant FKZ 50 OR 1101.  We thank W.-R,~Hamann for
comments and help with PoWR code.

{\it Facilities:} \facility{ CXO (LETG/ACIS)}, \facility{Suzaku}

%

\end{document}